\documentclass[fleqn,10pt]{wlscirep}

\title{Magnetic domain walls as broadband spin wave and elastic magnetisation wave emitters}

\author[1,*]{Rasmus B. Holl\"{a}nder}
\author[1]{Cai M\"{u}ller}
\author[1,+]{Jeffrey McCord}

\affil[1]{Institute for Materials Science, Kiel University, Kiel, 24143, Germany}

\affil[*]{rh@tf.uni-kiel.de}

\affil[+]{jmc@tf.uni-kiel.de}

\begin{abstract}
We report on the direct observation of spin wave and elastic wave emission from magnetic domain walls in ferromagnetic thin films. Driven by alternating homogeneous magnetic fields the magnetic domain walls act as coherent magnetisation wave sources. Directional and low damped elastic waves below and above the ferromagnetic resonance are excited. The wave vector of the magnetoelastically induced acoustic shear waves is linearly tuned by varying the excitation frequency. Domain wall emitted magnetostatic surface spin waves occur at higher frequencies, which characteristics are confirmed by micromagnetic simulations. The distinct modes of magnetisation wave excitation from micromagnetic objects are a general physical phenomenon relevant for dynamic magnetisation processes in structured magnetic films. Magnetic domain walls can act as reconfigurable antennas for spin wave and elastic wave generation with control of the wave orientation.
\end{abstract}
\begin{document}

\flushbottom
\maketitle

\thispagestyle{empty}

\section*{Introduction}
Relying on local excitation, spin waves offer the possibility to substitute modern day electronics by wave-computing \cite{Chumak2015,Kruglyak2010,Lenk2011}. Different kinds of applications like spin wave logic devices, signal processors, and devices involving spin wave mediated spin currents are projected. Spin wave excitation in magnetic films is generally based on small lateral antennas for the local excitation of spin waves. One approach towards application oriented spin wave technology focusses on the interaction between spin waves and naturally occurring magnetic microstructures at edges \cite{Mushenok2017, Lohman2018} and magnetic domain walls \cite{Whitehead2017}. In that context, symmetric Bloch type domain walls and their spin wave eigenmodes have long been studied \cite{Winter1961}. Theoretical predictions include numeric and analytical models in uniaxial \cite{Shimokhin1991} and cubic ferromagnetic materials \cite{Alekseev1999}. Beyond regular Bloch walls, in patterned magnetic thin films different types of magnetic domain walls form. Dynamic spin wave experiments involving magnetic textures range from N\'{e}el domain walls for mode localization and guidance of spin wave modes \cite{Truetzschler2016, Wagner2016} to magnetic domain walls as a delimiter for spin waves \cite{Pirro2015} to excited magnetic vortex cores \cite{Wintz2016}. Numerical studies suggest the utilisation of domain walls as directional spin wave emitters \cite{Roy2010,VandeWiele2016}. An evidence of such a behaviour was shown only once \cite{Mozooni2015}. In general, for magnetic thin films dynamic domain wall effects maximize at excitation frequencies around the specific domain wall resonance, which fundamentally is below the ferromagnetic resonance of the magnetic host material \cite{Yuan1992, Mozooni2014, Mullenix2014}.\\
The possible generation of elastic magnetisation waves from local elastic strains was investigated numerically\cite{Barra2017} by solving the Landau-Lifshitz-Gilbert equation together with the elastodynamic equations using finite element simulations. A piezoelectric element generates local dynamic strain which is transferred to a magnetostrictive material. The simulations predict low loss characteristics for the elastic waves leading to enhanced magnetisation wave propagation lengths as compared to conventional field generated spin waves. The elastic magnetisation wave \cite{Note2018} excitation from local piezoelectric antennas due to excitation of alternating strain induced anisotropy in magnetostrictive ferromagnetic layers was shown experimentally \cite{Cherepov2014}. Alternatively, it was suggested that magnetic domain walls in magnetostrictive materials may also radiate elastic waves from domain walls through magnetoelastic coupling of a moving or vibrating domain wall\cite{Lord1967}. Changes of magnetisation inside the domain wall are directly transferred to local modulation of strain through magnetostriction. Yet, the theoretical study was limited to the $\mathrm{MHz}$-regime. No direct experimental proof for the described modes of wave radiation are reported so far. \\
The inverse magnetoelastic behaviour, the coupling from elastic waves to magnetisation dynamics has also been investigated. It was shown that elastic waves can interact with magnetic textures, such as magnetic vortices \cite{Foerster2017}. In ~\cite{Weiler2011} it was shown that elastic waves induce precessional motion of magnetisation in a magnetostrictive thin film deposited on a piezoelectric substrate through strain mediation.\\
Here, we focus on the excitation of spin waves and elastic magnetisation waves from magnetic domain walls. Domain walls with nanometer core width offer the opportunity for local and reconfigurable spin wave devices, by flexible positioning of the domain walls. We prove the antenna free emission of magnetostatic spin waves and, in particular, elastic magnetisation waves in ferromagnetic films from asymmetric Bloch and N\'{e}el type domain walls \cite{Hubert2008} by direct time-resolved magneto-optical imaging in conjunction with complementary simulations. We show that domain walls offer an alternative and flexible excitation scheme for elastic magnetisation waves and spin waves. \\
\section*{Results}
\subsection*{Experimental evidence of magnetisation wave emission}
\begin{figure}
	\centering
	\includegraphics[width=8.5 cm]{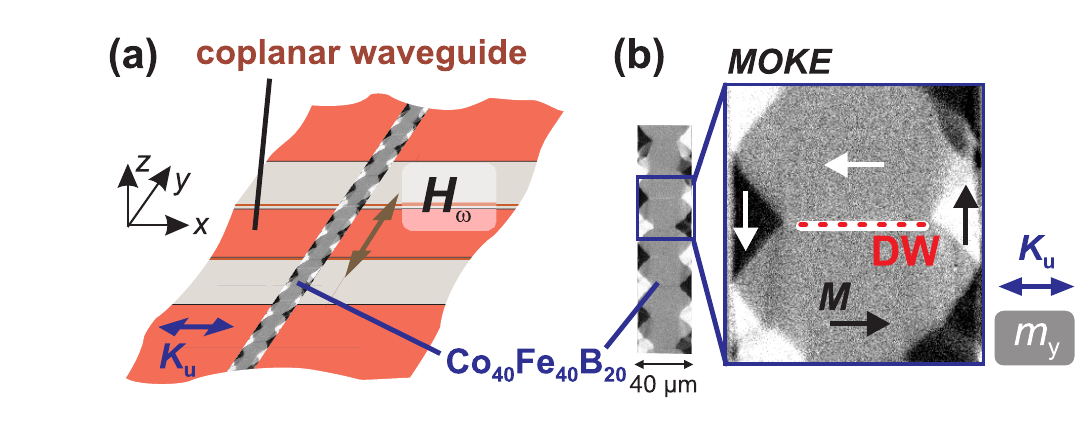}%
	\caption{Excitation scheme and experimental evidence. (a) Amorphous $\mathrm{Co_{40}Fe_{40}B_{20}}$ stripe with the easy axis of magnetisation $K_{\rm u}$ perpendicular to the long axis of the stripe excited by the homogeneous Oersted field on top of a coplanar waveguide with a 160~$\mathrm{\mu m}$ wide and 17.5~$\mathrm{\mu m}$ thick centre conductor. (b) Static ferromagnetic domain configuration after application of a saturating bias field along the $x$-axis, exhibiting a wide domain state (WDS). The position of the high angle domain wall (DW) is indicated.}
	\label{fig1}
\end{figure}
\begin{figure}
	\centering
	\includegraphics[width=8.0 cm]{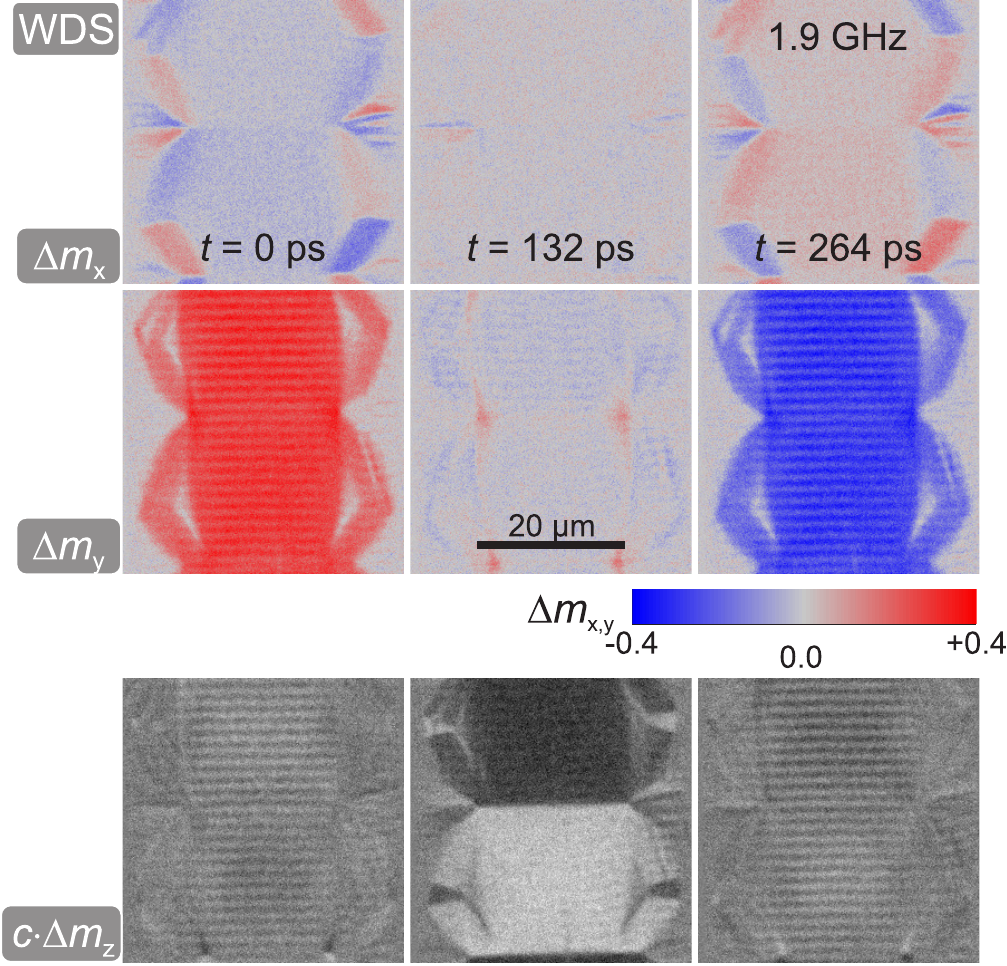}%
	\caption{Differential dynamic magnetisation response images along three different magneto-optical sensitivity directions at $t = 0~\mathrm{ps}$, $t = 132~\mathrm{ps}$, and $t = 264~\mathrm{ps}$ driven at the domain precessional frequency at $1.9~\mathrm{GHz}$ with an amplitude of $H_{\omega,\mathrm y} \approx 150~\mathrm{A/m}$.}
	\label{fig2}
\end{figure}
The investigated model system consists of a magnetostrictive amorphous ferromagnetic $\mathrm{Co_{40}Fe_{40}B_{20}}$ film with a thickness of $d_{\rm CoFeB} = 120~\mathrm{nm}$ deposited on a transparent glass substrate. To ensure the reproducible generation of domain walls the film is patterned into elongated stripes. Asymmetric $180^{\circ}$ Bloch walls form for this ferromagnetic layer thickness \cite{Hubert2008}. With the application of a magnetic bias field aligned perpendicular to the domain wall, the domain wall angle is reduced and asymmetric N\'{e}el walls form. For the standard experiment (Fig.~\ref{fig1}(a)), the axis of uniaxial anisotropy $K_{\mathrm u}$ is aligned perpendicular to the stripe axis along the $x$-direction. In Fig.~\ref{fig1}(b) a typical static remanent magnetisation state imaged by magneto-optical Kerr effect (MOKE) microscopy\cite{McCord2015} is shown, displaying a wide domain state (WDS). In the centre of the magnetic stripe, a periodic pattern of domains aligned parallel and antiparallel to the $x$-direction is obtained in the remanent state after applying a magnetic field along $x$. \\
Dynamic magnetisation component-selective MOKE response \cite{Hollaender2017} images are displayed in Fig.~\ref{fig2}. The differential time evolution of the individual magnetisation components $\Delta m_{\mathrm x}$, $\Delta m_{\mathrm y}$, and $\Delta m_{\mathrm z}$ are shown for three different times $t$. The excitation frequency of the sinusoidal varying field $H_{\omega,\mathrm y}$ was set to the domain resonance ${\omega}/{2\pi}=1.9~\mathrm{GHz}$. While $\Delta m_{\mathrm y}$ and $\Delta m_{\mathrm z}$ indicate a strong uniform precession of the domains around the easy axis of magnetisation, $\Delta m_{\mathrm x}$ shows only small deviations in time. The main features at the boundary to the closure domains visible in $\Delta m_{x}$ correspond to the nature of a magnetically charged domain state at the edges. In $\Delta m_{y}$ and $\Delta m_{\mathrm z}$ small periodic features appear inside the excited domains parallel to the domain wall. In the following, we focus on the magnetisation waves emitted by the central domain walls. As we show, these correspond to coherent elastic wave superpositions resulting from wave emission from the dynamically excited domain walls.\\
\begin{figure}
	\centering
	\includegraphics[width=18 cm]{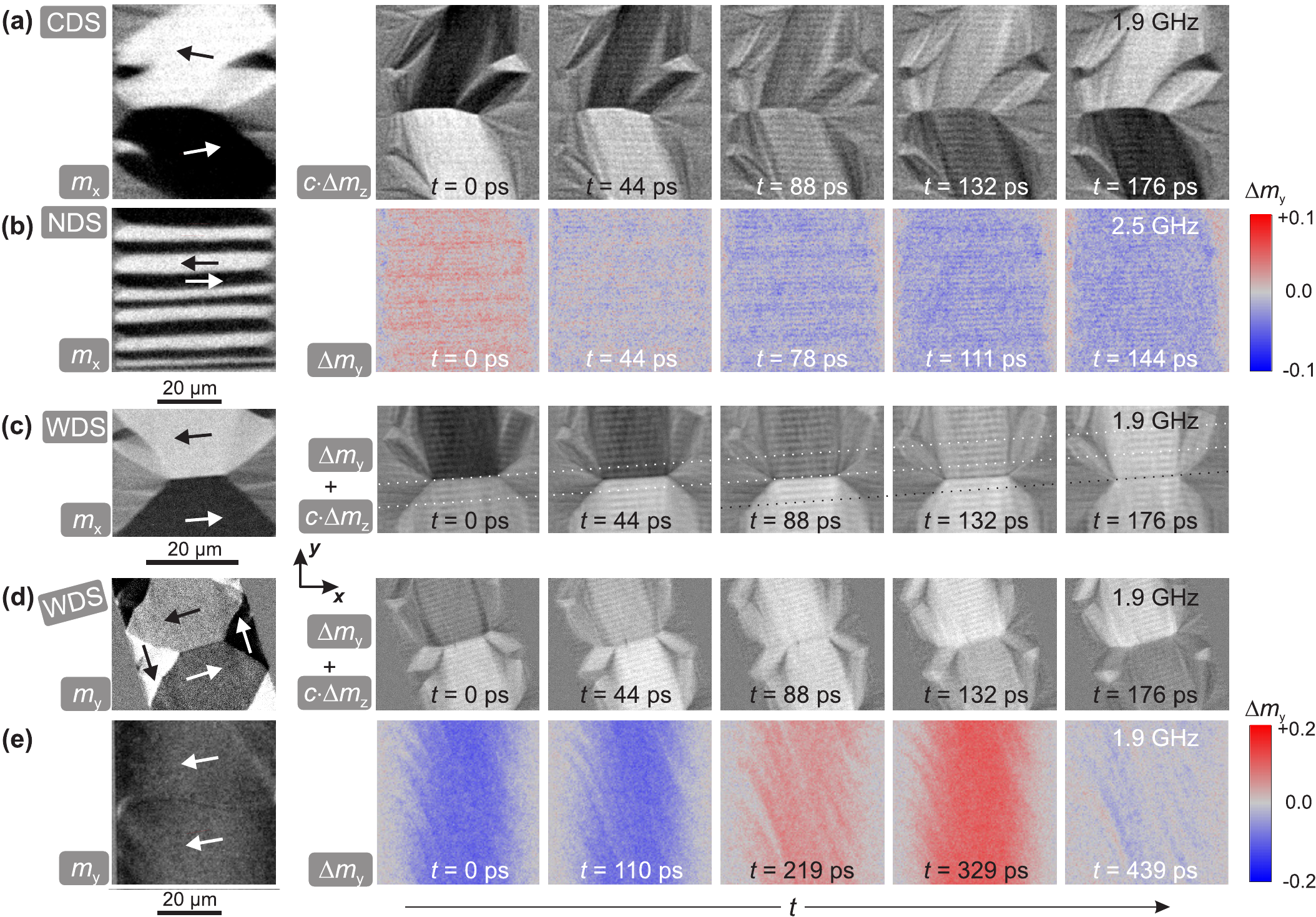}%
	\caption{(a) Static domain configuration and time evolution of a canted domain state (CDS) by application of a small bias field along the $y$-direction to a wide domain state (WDS). The static magnetisation component along the $x$-axis and the out-of-plane time evolution are shown. (b) A narrow domain state (NDS) at zero field created by field history perpendicular to uniaxial anisotropy. The static $x$-component of magnetisation is shown, while the time evolution shows pure longitudinal contrast along the $y$-direction. (c) Static and dynamic behaviour with tilted orientation of magnetisation and domain wall orientation in the WDS state and (d) with tilted sample orientation. Static and dynamic images are obtained with oblique plane of incidence. (e) Magnetic domain state close to saturation by applying a bias field parallel to the $x$-axis. The static component along the $y$-direction is depicted as well as the pure longitudinal time evolution along the $y$-direction. $H_{\omega \mathrm ,y} =150~\mathrm{A/m}$ was applied for all dynamic images. Principal directions of magnetisation are indicated in the static images.}
	\label{fig3}
\end{figure}
\subsection*{Domain wall dependent emission}
To prove the direct connection of the elastic wave generation to the magnetic domain walls we investigated the dependence of wave excitation on the domain and domain wall configuration. To ensure comparability each domain state was measured at its ferromagnetic resonance. Fig.~\ref{fig3}(a) shows the wave generation in a canted domain state (CDS). The CDS was created from the zero field WDS state by applying a bias field $H_y = 640~\mathrm{A/m} $ along the $y$-direction. The magnetisation in the centre domains is tilted by $10^{\circ}$ relative to $K_{\mathrm u}$ as indicated in Fig.~\ref{fig3}(a). Here, the domain walls transform from asymmetric Bloch to asymmetric N\'{e}el wall \cite{Hubert2008}, confirmed by MOKE imaging with the sensitivity aligned perpendicular to the domain wall (not shown). The time evolution of the out-of-plane MOKE contrast $c \cdot \Delta m_z$ displays wavefronts aligned parallel to the domain wall at the same excitation frequency of ${\omega}/{2\pi}=1.9~\mathrm{GHz}$ as used for the experiments shown in Fig.~\ref{fig2}. With the altered domain wall characteristics, the orientation and characteristics of the elastic waves remain. The wave vector does not follow the rotation of magnetisation.\\
By changing the domain width and field excitation frequency the elastic waves remain excited. Fig.~\ref{fig3}(b) shows a narrow domain state (NDS) generated by saturating the stripe along the $y$-direction and reducing the external field $H_{\mathrm y}$ to zero. The narrow domain width leads to different effective fields as compared to Fig.~\ref{fig1}(b), giving rise to a higher domain resonance frequency\cite{Queitsch2006,Hengst2014} of ${\omega}/{2\pi}=2.5~\mathrm{GHz}$. From the time evolution of $\Delta m_y$, it is directly apparent that the periodicity of the elastic waves shifted to smaller wavelengths. The static NDS shows various domain widths, leading to locally varying effective fields. The wavelength of the elastic wave is unaffected by the local effective fields.\\
A dependency on geometric factors like stripe orientation is excluded by changing the orientation of the domain walls relative to the excitation field and stripe axis. Fig.~\ref{fig3}(c) depicts a WDS with slightly tilted anisotropy. Statically the $m_x$ component is shown, while the time evolution was recorded with oblique plane of incidence resulting in a superposition of longitudinal and polar magneto-optical contrast ($\Delta m_y + c \cdot \Delta m_z$). Due to the alignment of the anisotropy the domain wall is oriented with an angle of $4.5^{\circ}$ to the $x$-axis. It is directly evident that the wave pattern follows the domain wall orientation and not the orientation of the magnetic track. In order to exclude the possibility of purely microwave field generated waves, an additional control experiment was performed, where the magnetic sample was turned with reference to the excitation source. Fig.~\ref{fig3}(d) displays these results for a WDS. Here, the static image displays the $m_y$ component of magnetisation and the time evolution of magnetisation changes shows a superposition of longitudinal and polar contrast ($\Delta m_y + c \cdot \Delta m_z$). The axis of oscillating magnetic field resided along the $y$-axis, while the long axis of the magnetic stripe was oriented $17^{\circ}$ compared to the $y$-axis.\\
Finally, no magnetization waves form in the absence of magnetic domain walls. An additional configuration without high-angle domain walls is depicted in Fig.~\ref{fig3}(e), where a bias field of $H_{\mathrm x} = 3~\mathrm{kA/m}$ was applied along the $x$-direction to obtain a nearly saturated domain state. In contrast to the examples before and in the absence of magnetic domain walls no elastic waves are excited. The experiments prove that the wave source is not the microwave field. The elastic wave generation and orientation is directly bound to the existence and orientation of the magnetic domain walls and independent of the alignment of magnetisation. \\
\begin{figure}
	\centering
	\includegraphics[width=15 cm]{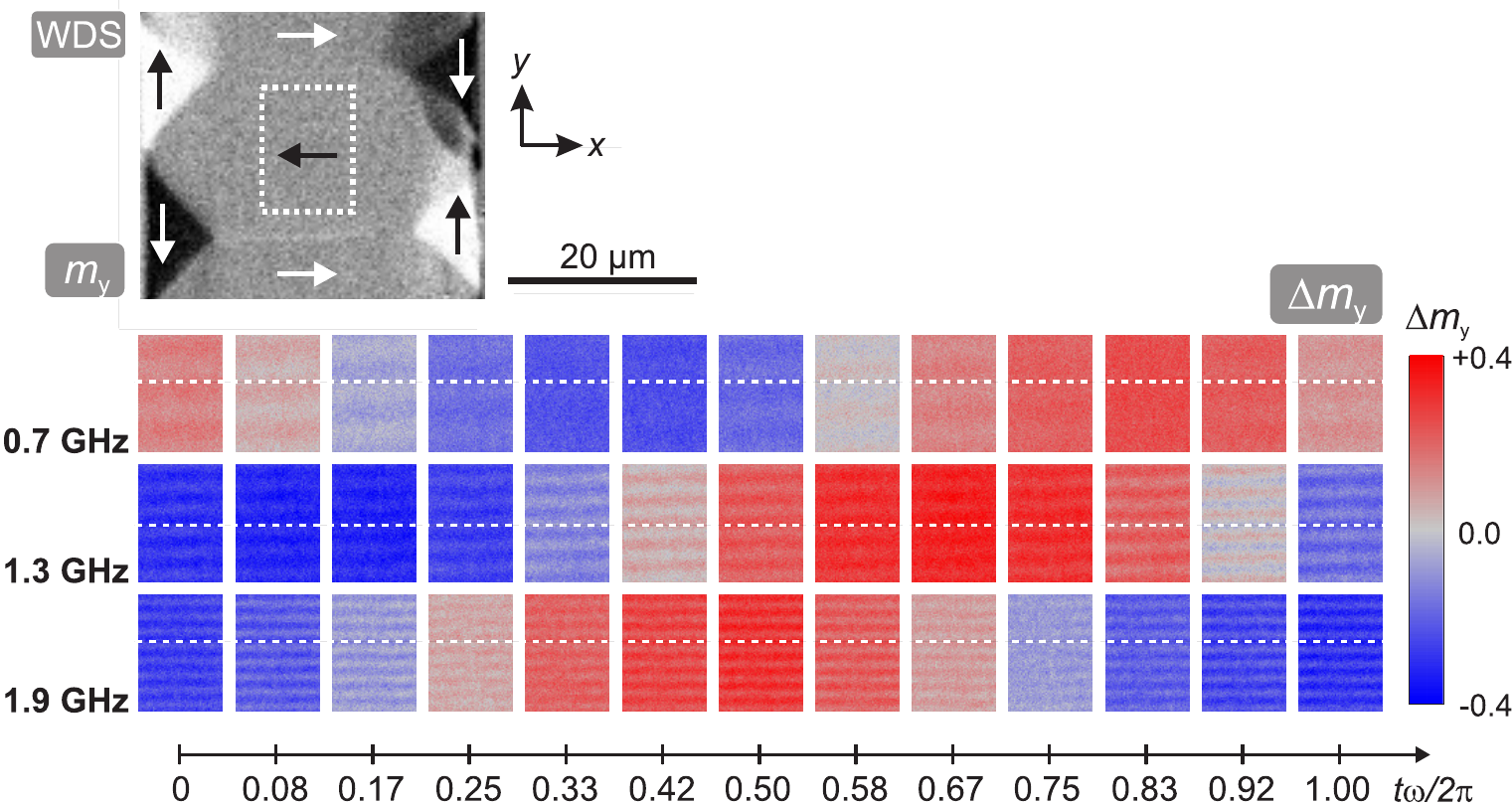}%
	\caption{Time evolution of an individual domain in the same WDS with the magneto-optical sensitivity along the $y$-direction  at different excitation frequencies of $0.7~\mathrm{GHz}$, $1.3~\mathrm{GHz}$, and $1.9~\mathrm{GHz}$ (see also Supplementary Movie 1). The analysed domain region is indicated in the static domain image. An Oersted field amplitude of $H_{\omega}=295~\mathrm{A/m}$ was used for the experiments. Lines for eye-guidance are drawn, starting at the node of the wavefront at $t\omega/2\pi = 0$.}
	\label{fig4}
\end{figure}

The ability to tune the emission of domain wall bound waves is demonstrated by measuring the elastic magnetisation wave characteristics in an identical WDS configuration with the excitation frequency ranging from ${\omega}/{2\pi}=0.7~\mathrm{GHz}$ to ${\omega}/{2\pi}=2.5~\mathrm{GHz}$. Exemplary results extracted from the centre domain at three different excitation frequencies are displayed in Fig.~\ref{fig4}. The resulting overall time evolution $\Delta m_y$ indicates the precessional magnetisation response of the magnetic domains. The change in wavelength of the elastic magnetisation waves, and thus, the dispersion relation of the excited coherent elastic waves are directly imaged. From the images it is evident that the wavelength of the superimposed elastic waves decreases with increasing excitation frequency. We interpret this as neighboring $180^{\circ}$-walls emitting waves with the same wavelength, which create standing wave characteristics inside the domains. For better clarity, reference lines parallel to the wavefronts, indicating the position of one node at $2 \pi t/\omega=0$, are drawn in Fig.~\ref{fig4}. In the case of ${\omega}/{2\pi} = 0.7~\mathrm{GHz}$ and ${\omega}/{2\pi} = 1.9~\mathrm{GHz}$ the reference lines remain on the node indicating standing wave characteristic. Only at ${\omega}/{2\pi} = 1.3~\mathrm{GHz}$ the elastic magnetisation waves exhibit small velocity propagating wave characteristics. In the obtained overall set of data, there appears to be no systematic correlation of gradual propagation or standing wave behaviour. This indicates a contribution from waves emitted at the closure domain walls, creating a spatial beat pattern due to slightly different wave emission periodicities and directions. Since the same unaltered domain and domain wall structure exhibits, both propagating and standing waves in Fig.~\ref{fig4}, an influence of the domain wall substructure is excluded. It should be noted that the magnetisation configuration cannot inhibit elastic waves from traveling across the magnetic boundaries formed by the magnetic domain walls.\\

\subsection*{Magnetostatic surface spin wave emission}
\begin{figure}
	\centering
	\includegraphics[width=8.5 cm]{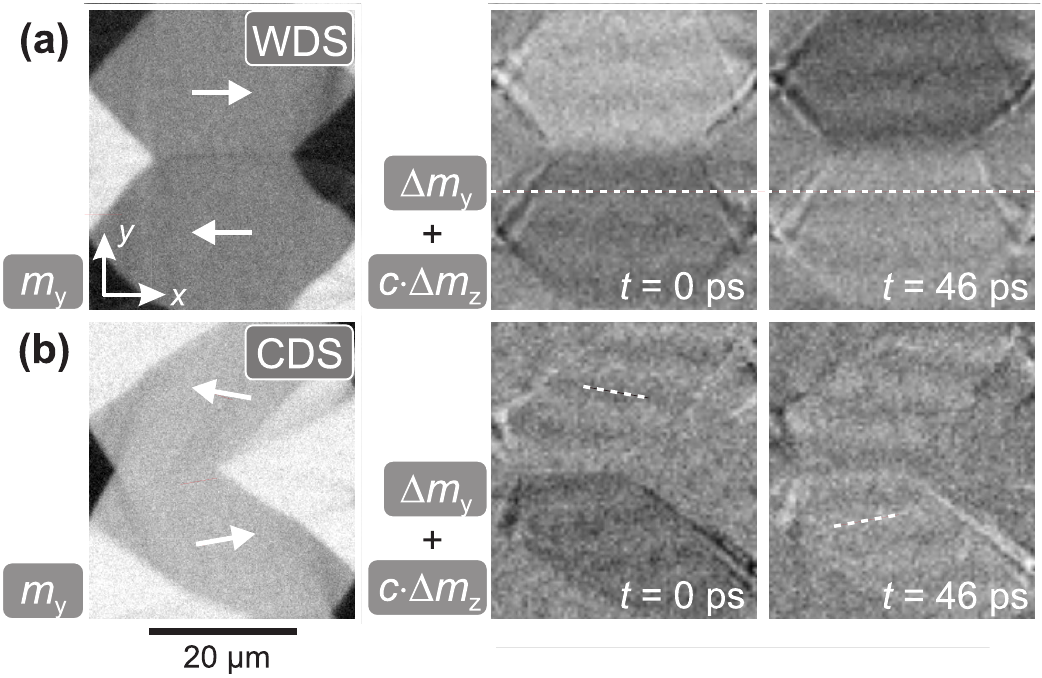}%
	\caption{Static configuration and time evolution of magnetisation at $9~\mathrm{GHz}$ excitation for (a) a remanent domain configuration (WDS) with field history along the axis of uniaxial anisotropy (see also Supplementary Movie 2 for $ c \cdot \Delta m_z$) and (b) a canted domain state (CDS) by application of a small bias field along the $y$-direction. Principal directions of magnetisation are indicated. The time evolution shows the qualitative magneto-optical magnetisation response as superposition of longitudinal and polar contrast. An excitation amplitude of $H_{\omega} \approx 100~\mathrm{{A}/{m}}$ was used for the shown time evolutions.  }
	\label{fig5}
\end{figure}
Beforehand, we showed that the elastic waves follow the domain wall orientation and not the orientation of magnetisation. Now, we compare our results to regular magnetostatic surface spin waves in the Damon-Eshbach-configuration (magnon wavector $\mathbf{k}$ and effective field $\mathbf{H}_{\mathrm{eff}}$ both in-plane and $\mathbf{H}_{\mathrm{eff}} \perp \mathbf{k}$). Two different states of magnetisation are compared in Fig.~\ref{fig5}. Fig.~\ref{fig5}(a) displays a remanent WDS domain configuration (as in Fig.~\ref{fig1}(b) and Fig.~\ref{fig4}). The exemplary images from the temporal response at ${\omega}/{2\pi} = 9~\mathrm{GHz}$ show the corresponding dynamic response. A line indicates the orientation of the standing spin wave nodes. Fig.~\ref{fig5}(b) shows a qualitative repetition of the experiment in Fig.~\ref{fig3}(a) (CDS), but for magnetostatic surface spin waves obtained at the excitation frequency of ${\omega}/{2\pi} = 9~\mathrm{GHz}$. It is directly evident that the detected spin waves, and in contrast to the elastic waves, now follow the orientation of magnetisation. The magnetostatic spin waves generated at increased frequency are not tied to the domain wall orientation but to the alignment of magnetisation.\\
\subsection*{Modelling of magnetisation wave emission}
\begin{figure}
	\centering
	\includegraphics[width= 8 cm]{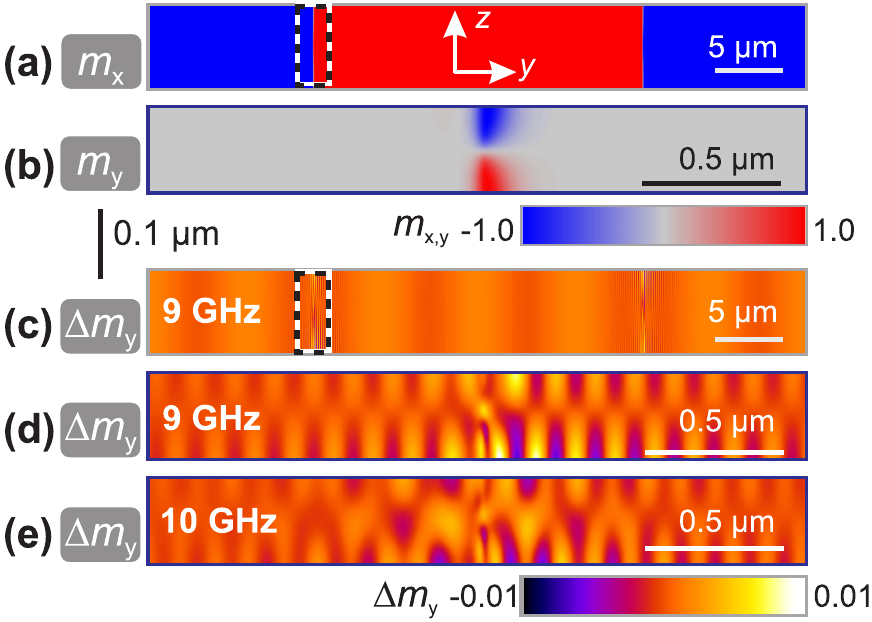}%
	\caption{Static and dynamic micromagnetic simulation of a periodic pattern consisting of two $180^{\circ}$ asymmetric Bloch walls separated by 25 $\rm \mu m$ wide domains. (a) Static cross-section in $m_x$ and (b) static cross-section in $m_y$ from the direct circumference of the left domain wall (as indicated in (a)). (c) Differential dynamic response along $\Delta m_y$ at $9~\mathrm{GHz}$, and local differential dynamic response $\Delta m_y$ at (d) $9~\mathrm{GHz}$ (see also Supplementary Movie 3) and (e) at $10~\mathrm{GHz}$.}
	\label{fig6}
\end{figure}
Micromagnetic simulations\cite{Vansteenkiste2014} were used to clarify the origin of occurring magnetisation waves. Selected simulation results are depicted in Fig.~\ref{fig6}. Fig.~\ref{fig6}(a) and (b) show the two dimensional static domain and domain wall configuration, where Fig.~\ref{fig6}(a) displays the simulated full cross-section as $m_x$ and Fig.~\ref{fig6}(b) the close circumference of one $180^{\circ}$ asymmetric Bloch wall as $m_y$. Fig.~\ref{fig6}(c) shows the differential magnetisation response along the $y$-direction at an excitation frequency of $9~\mathrm{GHz}$. Standing magnetostatic surface spin waves are visible inside the domains. The waves are in phase on both surfaces. Apart from the standing magnetostatic surface spin waves in the simulations, the domain walls emit short wavelength ($\lambda<300~\mathrm{nm}$) propagating spin waves at excitation frequencies exceeding $3~\mathrm{GHz}$. Examples are depicted in Fig.~\ref{fig6}(d) and (e) for a frequency of $9~\mathrm{GHz}$ and $10~\mathrm{GHz}$, respectively. As the wavelengths of these spin waves are below the optical limit of detection, they do not have an experimental counterpart in this work. At frequencies below $5~\mathrm{GHz}$ the low wavelength spin waves are found to be oscillating at higher harmonics of the excitation frequency. At frequencies exceeding $5~\mathrm{GHz}$ this first low wavelength mode is excited directly at the excitation frequency. Furthermore, a second low wavelength mode is emitted at frequencies equal or higher than $10~\mathrm{GHz}$, as evident by comparison of Fig.~\ref{fig6}(d) and (e). The data presented in Fig.~\ref{fig6}(c) to (e) was simulated in dynamic equilibrium.\\
\begin{figure}
	\centering
	\includegraphics[width= 7 cm]{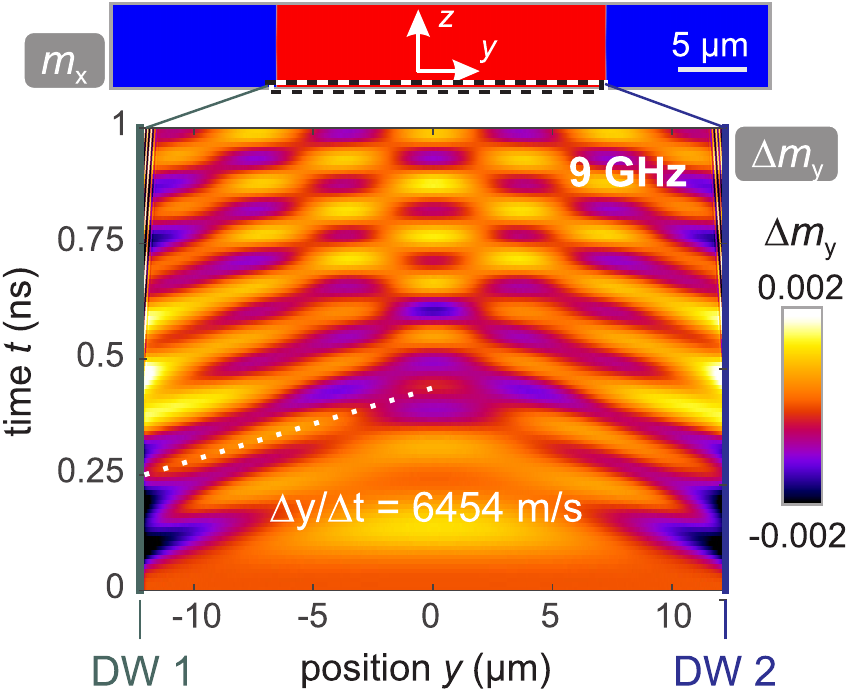}%
	\caption{Dynamic micromagnetic simulation of the transient effect in the top layer ($z=120~\mathrm{nm}$) and central domain (red in Fig.~\ref{fig6}(a)) at an excitation frequency of $9~\mathrm{GHz}$. The position of the two domain walls (DW 1 and DW 2) is indicated at the left and right axes. The phase velocity of the MMSW can be determined by following the wave profile along the propagation direction.}
	\label{fig7}
\end{figure}
In order to determine the location where the magnetostatic surface spin waves originate, we simulated the transient effect. The corresponding data for an excitation frequency of $9~\mathrm{GHz}$ (compare to Fig.\ref{fig6}(c)) is presented in Fig.~\ref{fig7}. Here, the evolution of dynamic magnetisation response $\Delta m_y$ is shown for the bottom layer normal to the $z$-axis and in the central domain of the simulated pattern (compare to Fig.~\ref{fig6}(a)). At $t=0~\mathrm{ns}$ the application of the sinusoidal microwave field was started. No wavefronts appear in the domain at the beginning. With increasing time, wavefronts from the domain walls propagate into the central domain and start to form superposition patterns. By following the phase of one wavefront, the phase velocity can be obtained to be $v_{\mathrm{9 GHz}}=\Delta y / \Delta t =6454~\mathrm{m/s}$. Our calculations prove that the spin waves originate from domain walls emitting spin waves with the same wavelength. The underlying mechanism of spin wave emission from domain walls can be understood by means of local effective fields present in the domain wall \cite{Schloemann1964}.\\
\begin{figure}
	\centering
	\includegraphics[width= 8 cm]{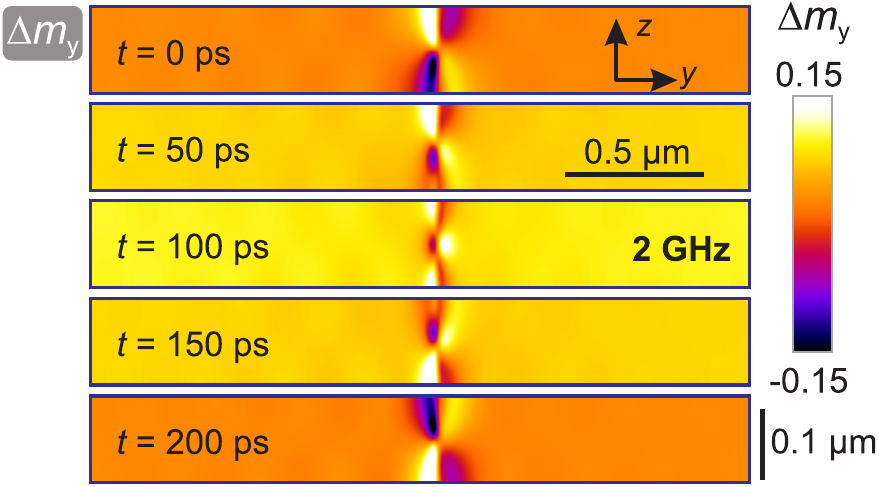}%
	\caption{Dynamic micromagnetic simulation of the time evolution of Fig.~\ref{fig6}~(b) at an excitation frequency of $2~\mathrm{GHz}$.}
	\label{fig8}
\end{figure}
Yet, the elastic magnetisation wave branch is not reproduced in the numerical computer experiments, confirming an alternative mechanism of magnetisation wave generation. Fig.~\ref{fig8} displays the results obtained by micromagnetic simulation at a low $2~\mathrm{GHz}$ excitation frequency in dynamic equilibrium. Only the close circumference of the domain wall is shown. The precession of the two domains around the domain wall can be seen directly by the change of magnetization $\Delta m_{y}$ of the domains with time. Furthermore, an even higher activity $\Delta m_{y}$ is experienced at the domain walls. The dynamic tensioning inside the domain wall, via magnetoelastic coupling, generates elastic waves. The changing domain wall acts as a localized alternating body force inside the elastic material from which elastic waves penetrate the material. In the region of the domain wall the magnetisation rotates by $\pi$, meaning that there is at least one point in this region where the lateral magnetostriction curve has the highest possible slope. At these points the highest dependence of the dynamic strain on the varying magnetisation is expected. Magnetoelastic coupling leads to a coherent dependence of the elastic response on the dynamic magnetisation response in the domain wall. The dispersion relation of the elastic waves provides proof for the domain wall mediated generation of elastic waves.
\subsection*{Dispersion of distinct mode emission}
\begin{figure}
	\centering
	\includegraphics[width= 8 cm]{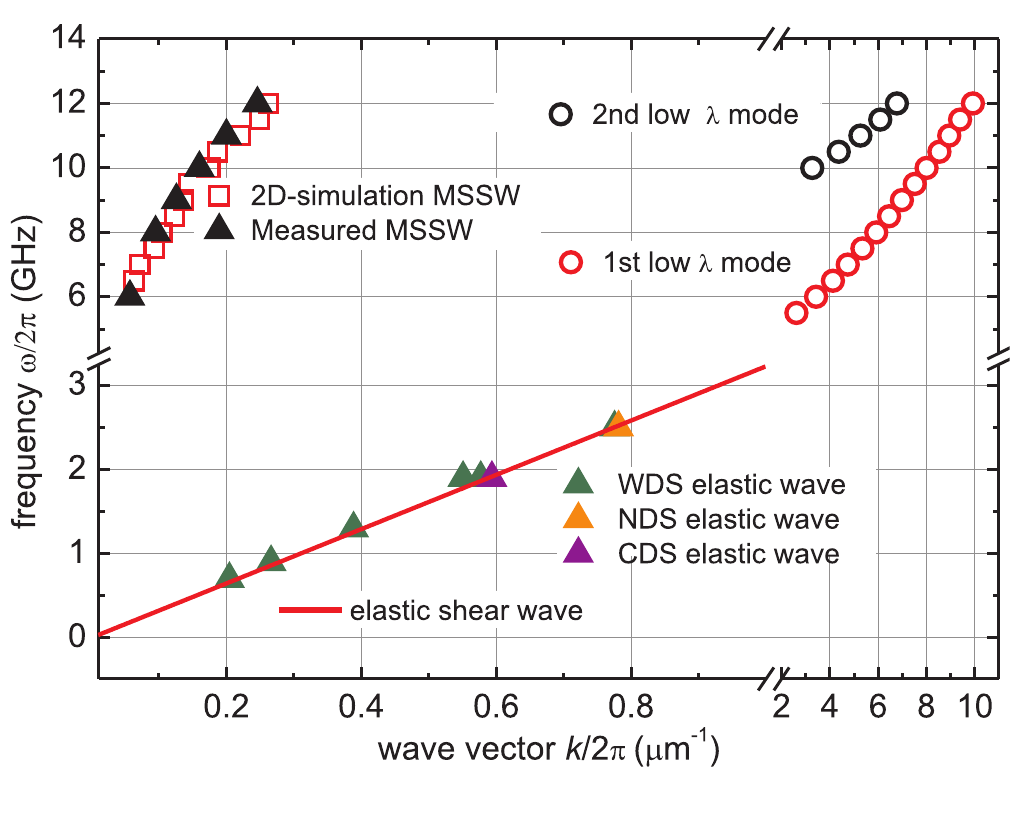}%
	\caption{Wave vector to excitation frequency dependence of emerging wave-fronts parallel to the $x$-axis for elastic waves and magnetostatic surface spin wave (MSSW) modes together with numerical data on the magnetostatic surface spin waves and the linear dispersion of an elastic shear wave in $\mathrm{Co_{40}Fe_{40}B_{20}}$.}
	\label{fig9}
\end{figure}
By analysing the MOKE images inside a central domain by averaging along the wavefront and subsequent fast Fourier transformation in time and space, the characteristic dependencies between excitation frequency $\omega$ and wave vector $k$ of the emitted spin waves are extracted. The results are directly compared to simulation results. A complete analysis of the magnetisation wave dispersion from dynamic imaging and micromagnetic simulation is displayed in Fig.~\ref{fig9}.\\
The experimentally found magnetostatic spin wave dispersion is reproduced by the micromagnetic simulations. Furthermore, a parabolic dispersion, a signature of magnetostatic surface spin waves, is found at frequencies ranging from 6 to 12~$\mathrm{GHz}$. In the same frequency range, in the vicin ity of the DWs two additional low wavelength modes are found in the simulations. No experimental evidence for such spin waves is found as the smaller wavelength spin wave modes are not accessible by MOKE imaging due to the limited spatial resolution of the optical experiment. The regime of elastic (spin) wave generation is not confirmed by the purely micromagnetic modelling.\\
Experimentally, the elastic magnetisation wave branch is not affected by small changes of the domain configuration. Elastic waves are observed for frequencies ranging from $0.55~\mathrm{GHz}$ up to $2.5~\mathrm{GHz}$, the upper bound limited by the spatial optical resolution. The extracted $\omega$-$k$ dependence reveals that WDS, NDS, and CDS share a single linear dispersion relation. The dispersion is independent of the state of magnetisation. To conclusively confirm a magnetoelastic mechanism of elastic magnetisation wave generation, the dispersion relation is compared to the expected linear dependence for an elastic shear wave. For this the velocity of sound $v$ was calculated using 
\begin{equation}\label{shear-wave}
v=\sqrt{\frac{G}{\rho}},
\end{equation}
where $G=70~\mathrm{GPa}$ is the shear modulus of CoFeB \cite{Krawczyk2017}. With the volumetric mass density\cite{Krawczyk2017} $\rho=7050~ \mathrm{{kg}/{m^3}}$, a shear wave propagation velocity of $v=3.151~\mathrm{{km}/{s}}$ is calculated. The corresponding dispersion relation is added to Fig.~\ref{fig9}. It perfectly coincides with the dispersion of the experimentally found elastic wave modes. No adjustment of parameters was performed. The domain wall tied dispersion relation derived from the dynamic MOKE images corresponds to a sound velocity of  $v=3.159~\mathrm{{km}/{s}}$. This nearly perfect agreement finally confirms the presumed elastic shear wave mechanism. 
\section*{Conclusion}
Distinct modes of magnetisation waves are generated from domain wall dynamics. The existence of the waves is confirmed by direct time resolved magneto-optical imaging. Standing magnetostatic surface spin waves in the Damon-Eshbach configuration and elastic magnetisation waves are excited by magnetic domain walls acting as antennas. The experimental investigation shows that the magnetostatic surface spin waves follow the alignment of magnetisation orientation in the domains. In contrast, the elastic waves are independent of the alignment of the excitation field and the state of magnetisation inside the domains. The orientation of the elastic waves is purely bound to the orientation of the domain walls. They are solely and directly connected to the dynamically excited domain walls. The direction of wave propagation can be tuned with the orientation of the domain walls.\\ 
Together with the experimental evidence, our analysis of the linear dispersion relation clearly shows that the origin of the elastic mode stems from coherent elastic waves generated at the domain walls in the magnetostrictive thin films through magnetoelastic coupling. Origin is a dynamic tensioning inside the excited domain walls of the magnetostrictive material. \\
The discovery of the alternative mode of acoustic wave generation from domain walls could help building broadband and reconfigurable sources of low damped magnetoelastic spin waves for future applications without the need for piezoelectric substrates or elements. The magnetostatic and elastic magnetisation waves emitted by high angle domain walls show that magnetic domain walls can be used as reconfigurable sources for coherent emission of magnetisation waves up to high frequencies into the $\mathrm{GHz}$-regime. \\
%

\section*{Methods}
%
\subsection*{Magnetostrictive CoFeB thin film stripe arrays}
A $\rm Ta (5~nm)/Co_{40}Fe_{40}B_{20} (120~nm)/Ru (3~nm)$ layer was prepared by sputter deposition on a glass $\rm SiO_{2}$ wafer with a thickness $t =(800~\rm \mu m)\rm$. The Ru covering layer forms a protective layer over the ferromagnetic structures to avoid oxidation and corrosion. Saturation polarisation of the ferromagnetic material is $J\rm _{s}=1.5T$. An induced uniaxial magnetic anisotropy of $K_{\rm u}\rm =1300~J/m^{3}$ is introduced by applying a magnetic field of $H_{\rm dep}\rm=8~kA/m$ during magnetic layer deposition \cite{Hengst2014}. The saturation magnetostriction is $\lambda_{\mathrm s} \approx 23\times 10^{-6}$ \cite{OHandley2000}. To ensure the generation of periodic domain walls the film is patterned into elongated stripes with the dimension of $40~\mathrm{\mu m}~\times~10~\mathrm{mm}$ via standard photolithography and selective ion beam etching. Substrate pieces of $\rm 10~mm \times 10~mm$ were cut by wafer sawing from the full wafer.
\subsection*{Dynamic magnetisation characterisation}

The dynamic magnetic characteristics, e.g. magnetic permeability spectra, were obtained by pulsed inductive microwave magnetometry (PIMM) \cite{Silva1999} with an in-plane magnetic field pulse of $H_{pulse}\approx\rm 3.0{A/m}$ and a rise time of $t_{rt,10-90}\leq50\rm ~ps$ (see Supplementary Figure 1). The measurements were performed with varying bias field $H_{ext}$. The time domain data was transferred into the frequency domain by fast Fourier transformation, from which the dynamic magnetic permeability spectra were obtained. From the experiments the individual domain resonance in the stripes and the effective damping parameter $\alpha=0.008$ of the material were obtained.
\subsection*{Component selective time resolved magneto-optical wide-field imaging}
To experimentally probe wave emission, we apply time-resolved magneto-optical wide field imaging \cite{McCord2015} based on the magneto-optical Kerr effect (MOKE) to a ferromagnetic thin film structure. For all magneto-optical images a specialized magneto-optical wide-field microscope is used that allows not only for static domain imaging, but also for time-resolved observations by using a pulsed $\mathrm{Nd:YVO_4}$ laser illumination system with $7~\mathrm{ps}$ pulse width and a repetition rate of $50~\mathrm{MHz}$ for stroboscopic imaging. A motorised stage is implemented in the setup allowing for precise and reproducible positioning of a lens, resulting in automatic control of the angle of incidence and magneto-optical sensitivity. A coplanar waveguide is utilised to generate a homogeneous high frequency magnetic field $H_{\omega\mathrm ,y}$. The central wave guide consists of a 160~$\mathrm{\mu m}$ wide and 17.5~$\mathrm{\mu m}$ thick centre conductor. The sample is positioned with the magnetic surface down on top of the waveguide and the sample direction relative to the dynamic magnetic field can be freely adjusted. To influence the domain pattern, and thus the domain wall determined magnetisation dynamics, the magnetic field history is varied in angle and strength.\\
The dynamic change of individual magnetisation components is probed by a component selective imaging process\cite{Hollaender2017}. It allows for direct measurement of in-plane and out-of-plane magnetisation components by dynamic MOKE microscopy. In order to measure only dynamic magnetisation contrast each dynamic MOKE image at a phase $\phi$ is subtracted by an image of phase $\phi-\pi$. The measurement process involves the measurement using two non-zero angles of incidence differing by a sign change and subtraction of those two magneto-optical sensitivities. By following this procedure the polar magneto-optical contrast cancels out, leading to pure magnetic in-plane response. By this also a possible birefringence or photo-elastic effect resulting from the magnetoelastic interaction cancels for the detection of the pure in-plane magnetisation components, as the reflection coefficients are symmetric for two opposing angles of incidence \cite{Silverman1990}. Only magnetic contrast changes are visible in the quantitative dynamic in-plane images ($\Delta m_{\mathrm x}$, $\Delta m_{\mathrm y}$). 

\subsection*{Micromagnetic modelling}
Two-dimensional micromagnetic simulations using mumax${}^3$~\cite{Vansteenkiste2014} were used to explain the experimental results. In accordance with the experiments, in the numerical simulations a two-dimensional two domain state of domain width $w=25~\mathrm{\mu m}$  was simulated in the $y,z$ cross-section plane. Periodic boundary conditions along $x$- and $y$-directions were applied. A grid size of $1 \times 12500 \times 30$ with a cell size of $4 \times 4 \times 4~\mathrm{nm^3}$ was used. Calculations were performed with an exchange constant $A = 15~\mathrm{pJ/m}$ \cite{Conca2013}, a damping constant $\alpha=0.008$, a uniaxial anisotropy ${K_\mathrm u}=1300~\mathrm{J/m^3}$ and at temperature $T=0~K$. The domain wall was excited with a magnetic field amplitude of $H_{\omega \mathrm ,y}=145~\mathrm{{A}/{m}}$. The homogeneous high frequency field was applied at frequencies ranging from $0.5~\mathrm{GHz}$ to $12~\mathrm{GHz}$. In order to gain comparable results and gain purely dynamic magnetisation response $\Delta m_y$ each simulated time frame at excitation phase $\phi$ is subtracted by the corresponding time frame at excitation phase $\phi-\pi$.

\section*{Acknowledgements}
Funding through the German Research Council (DFG, MC9/10-2) is highly appreciated. J.M. thanks the DFG for support through their Heisenberg Programme (DFG, MC9/9-1, MC9/9-2). We thank Roland Mattheis for thin film preparation and Ingolf M\"{o}nch for help with the sample preparation and Mathis Lohman for assistance with the initial imaging experiments.


\begin{thebibliography}{10}
	\expandafter\ifx\csname url\endcsname\relax
	\def\url#1{\texttt{#1}}\fi
	\expandafter\ifx\csname urlprefix\endcsname\relax\def\urlprefix{URL }\fi
	\expandafter\ifx\csname doiprefix\endcsname\relax\def\doiprefix{DOI }\fi
	\providecommand{\bibinfo}[2]{#2}
	\providecommand{\eprint}[2][]{\url{#2}}
	
	\bibitem{Chumak2015}
	\bibinfo{author}{Chumak, A.~V.}, \bibinfo{author}{Vasyuchka, V.~I.},
	\bibinfo{author}{Serga, A.~A.} \& \bibinfo{author}{Hillebrands, B.}
	\newblock \bibinfo{journal}{\bibinfo{title}{Magnon spintronics}}.
	\newblock {\emph{\JournalTitle{Nature Physics}}} \textbf{\bibinfo{volume}{11}},
	\bibinfo{pages}{453--461} (\bibinfo{year}{2015}).
	
	\bibitem{Kruglyak2010}
	\bibinfo{author}{Kruglyak, V.}, \bibinfo{author}{Demokritov, S.} \&
	\bibinfo{author}{Grundler, D.}
	\newblock \bibinfo{journal}{\bibinfo{title}{Magnonics}}.
	\newblock {\emph{\JournalTitle{Journal of Physics D: Applied Physics}}}
	\textbf{\bibinfo{volume}{43}}, \bibinfo{pages}{264001}
	(\bibinfo{year}{2010}).
	
	\bibitem{Lenk2011}
	\bibinfo{author}{Lenk, B.}, \bibinfo{author}{Ulrichs, H.},
	\bibinfo{author}{Garbs, F.} \& \bibinfo{author}{M{\"u}nzenberg, M.}
	\newblock \bibinfo{journal}{\bibinfo{title}{The building blocks of magnonics}}.
	\newblock {\emph{\JournalTitle{Physics Reports}}}
	\textbf{\bibinfo{volume}{507}}, \bibinfo{pages}{107--136}
	(\bibinfo{year}{2011}).
	
	\bibitem{Mushenok2017}
	\bibinfo{author}{Mushenok, F.~B.} \emph{et~al.}
	\newblock \bibinfo{journal}{\bibinfo{title}{Broadband conversion of microwaves
			into propagating spin waves in patterned magnetic structures}}.
	\newblock {\emph{\JournalTitle{Applied Physics Letters}}}
	\textbf{\bibinfo{volume}{111}}, \bibinfo{pages}{042404}
	(\bibinfo{year}{2017}).
	
	\bibitem{Lohman2018}
	\bibinfo{author}{Lohman, M.}, \bibinfo{author}{Mozooni, B.} \&
	\bibinfo{author}{McCord, J.}
	\newblock \bibinfo{journal}{\bibinfo{title}{Homogeneous microwave field emitted
			propagating spin waves: Direct imaging and modeling}}.
	\newblock {\emph{\JournalTitle{Journal of Magnetism and Magnetic Materials}}}
	\textbf{\bibinfo{volume}{450}}, \bibinfo{pages}{7 -- 12}
	(\bibinfo{year}{2018}).
	
	\bibitem{Whitehead2017}
	\bibinfo{author}{Whitehead, N.~J.}, \bibinfo{author}{Horsley, S. A.~R.},
	\bibinfo{author}{Philbin, T.~G.}, \bibinfo{author}{Kuchko, A.~N.} \&
	\bibinfo{author}{Kruglyak, V.~V.}
	\newblock \bibinfo{journal}{\bibinfo{title}{Theory of linear spin wave emission
			from a {Bloch} domain wall}}.
	\newblock {\emph{\JournalTitle{Physical Review B}}}
	\textbf{\bibinfo{volume}{96}}, \bibinfo{pages}{064415}
	(\bibinfo{year}{2017}).
	
	\bibitem{Winter1961}
	\bibinfo{author}{Winter, J.~M.}
	\newblock \bibinfo{journal}{\bibinfo{title}{{Bloch} wall excitation.
			application to nuclear resonance in a bloch wall}}.
	\newblock {\emph{\JournalTitle{Physical Review}}}
	\textbf{\bibinfo{volume}{124}}, \bibinfo{pages}{452--459}
	(\bibinfo{year}{1961}).
	
	\bibitem{Shimokhin1991}
	\bibinfo{author}{Shimokhin, I.~A.}
	\newblock \bibinfo{journal}{\bibinfo{title}{On the {Gilinskii} branch of the
			spectrum of excitations of a domain wall in uniaxial ferromagnetics}}.
	\newblock {\emph{\JournalTitle{physica status solidi (b)}}}
	\textbf{\bibinfo{volume}{167}}, \bibinfo{pages}{243--250}
	(\bibinfo{year}{1991}).
	
	\bibitem{Alekseev1999}
	\bibinfo{author}{Alekseev, A.~M.}, \bibinfo{author}{D{\"o}tsch, H.},
	\bibinfo{author}{Kulagin, N.~E.}, \bibinfo{author}{Popkov, A.~F.} \&
	\bibinfo{author}{Synogach, V.~T.}
	\newblock \bibinfo{journal}{\bibinfo{title}{Microwave excitations of a domain
			wall in a cubic magnet with induced anisotropy}}.
	\newblock {\emph{\JournalTitle{Technical Physics}}}
	\textbf{\bibinfo{volume}{44}}, \bibinfo{pages}{657--663}
	(\bibinfo{year}{1999}).
	
	\bibitem{Truetzschler2016}
	\bibinfo{author}{Tr\"{u}tzschler, J.}, \bibinfo{author}{Sentosun, K.},
	\bibinfo{author}{Mozooni, B.}, \bibinfo{author}{Mattheis, R.} \&
	\bibinfo{author}{McCord, J.}
	\newblock \bibinfo{journal}{\bibinfo{title}{Magnetic domain wall gratings for
			magnetization reversal tuning and confined dynamic mode localization}}.
	\newblock {\emph{\JournalTitle{Scientific Reports}}}
	\textbf{\bibinfo{volume}{6}}, \bibinfo{pages}{30761} (\bibinfo{year}{2016}).
	
	\bibitem{Wagner2016}
	\bibinfo{author}{Wagner, K.} \emph{et~al.}
	\newblock \bibinfo{journal}{\bibinfo{title}{Magnetic domain walls as
			reconfigurable spin-wave nanochannels}}.
	\newblock {\emph{\JournalTitle{Nature Nanotechnology}}}
	\textbf{\bibinfo{volume}{11}}, \bibinfo{pages}{432--436}
	(\bibinfo{year}{2016}).
	
	\bibitem{Pirro2015}
	\bibinfo{author}{Pirro, P.} \emph{et~al.}
	\newblock \bibinfo{journal}{\bibinfo{title}{Experimental observation of the
			interaction of propagating spin waves with n{\'e}el domain walls in a
			{Landau} domain structure}}.
	\newblock {\emph{\JournalTitle{Applied Physics Letters}}}
	\textbf{\bibinfo{volume}{106}}, \bibinfo{pages}{232405}
	(\bibinfo{year}{2015}).
	
	\bibitem{Wintz2016}
	\bibinfo{author}{Wintz, S.} \emph{et~al.}
	\newblock \bibinfo{journal}{\bibinfo{title}{Magnetic vortex cores as tunable
			spin-wave emitters}}.
	\newblock {\emph{\JournalTitle{Nature Nanotechnology}}}
	\textbf{\bibinfo{volume}{11}}, \bibinfo{pages}{948--953}
	(\bibinfo{year}{2016}).
	
	\bibitem{Roy2010}
	\bibinfo{author}{Roy, P.~E.}, \bibinfo{author}{Trypiniotis, T.} \&
	\bibinfo{author}{Barnes, C. H.~W.}
	\newblock \bibinfo{journal}{\bibinfo{title}{Micromagnetic simulations of
			spin-wave normal modes and the resonant field-driven magnetization dynamics
			of a $360\ifmmode^\circ\else\textdegree\fi{}$ domain wall in a soft magnetic
			stripe}}.
	\newblock {\emph{\JournalTitle{Physical Review B}}}
	\textbf{\bibinfo{volume}{82}}, \bibinfo{pages}{134411}
	(\bibinfo{year}{2010}).
	
	\bibitem{VandeWiele2016}
	\bibinfo{author}{Van~de Wiele, B.}, \bibinfo{author}{H\"{a}m\"{a}l\"{a}inen,
		S.~J.}, \bibinfo{author}{Bal\'{a}\^{z}, P.}, \bibinfo{author}{Montoncello,
		F.} \& \bibinfo{author}{van Dijken, S.}
	\newblock \bibinfo{journal}{\bibinfo{title}{Tunable short-wavelength spin wave
			excitation from pinned magnetic domain walls}}.
	\newblock {\emph{\JournalTitle{Scientific Reports}}}
	\textbf{\bibinfo{volume}{6}}, \bibinfo{pages}{21330} (\bibinfo{year}{2016}).
	
	\bibitem{Mozooni2015}
	\bibinfo{author}{Mozooni, B.} \& \bibinfo{author}{McCord, J.}
	\newblock \bibinfo{journal}{\bibinfo{title}{Direct observation of closure
			domain wall mediated spin waves}}.
	\newblock {\emph{\JournalTitle{Applied Physics Letters}}}
	\textbf{\bibinfo{volume}{107}}, \bibinfo{pages}{042402}
	(\bibinfo{year}{2015}).
	
	\bibitem{Yuan1992}
	\bibinfo{author}{Yuan, S.~W.} \& \bibinfo{author}{Bertram, H.~N.}
	\newblock \bibinfo{journal}{\bibinfo{title}{Magnetic thin film domain wall
			motion under dynamic fields}}.
	\newblock {\emph{\JournalTitle{Journal of Applied Physics}}}
	\textbf{\bibinfo{volume}{72}}, \bibinfo{pages}{1033--1038}
	(\bibinfo{year}{1992}).
	
	\bibitem{Mozooni2014}
	\bibinfo{author}{Mozooni, B.}, \bibinfo{author}{von Hofe, T.} \&
	\bibinfo{author}{McCord, J.}
	\newblock \bibinfo{journal}{\bibinfo{title}{Picosecond wide-field
			magneto-optical imaging of magnetization dynamics of amorphous film
			elements}}.
	\newblock {\emph{\JournalTitle{Physical Review B}}}
	\textbf{\bibinfo{volume}{90}}, \bibinfo{pages}{054410}
	(\bibinfo{year}{2014}).
	
	\bibitem{Mullenix2014}
	\bibinfo{author}{Mullenix, J.}, \bibinfo{author}{El-Ghazaly, A.},
	\bibinfo{author}{Lee, D.~W.}, \bibinfo{author}{Wang, S.~X.} \&
	\bibinfo{author}{White, R.~M.}
	\newblock \bibinfo{journal}{\bibinfo{title}{Spin-wave resonances in the
			presence of a {Bloch} wall}}.
	\newblock {\emph{\JournalTitle{Physical Review B}}}
	\textbf{\bibinfo{volume}{89}}, \bibinfo{pages}{224406}
	(\bibinfo{year}{2014}).
	
	\bibitem{Barra2017}
	\bibinfo{author}{Barra, A.}, \bibinfo{author}{Mal, A.},
	\bibinfo{author}{Carman, G.} \& \bibinfo{author}{Sepulveda, A.}
	\newblock \bibinfo{journal}{\bibinfo{title}{Voltage induced mechanical/spin
			wave propagation over long distances}}.
	\newblock {\emph{\JournalTitle{Applied Physics Letters}}}
	\textbf{\bibinfo{volume}{110}}, \bibinfo{pages}{072401}
	(\bibinfo{year}{2017}).
	
	\bibitem{Note2018}
	\bibinfo{note}{In literature the exhibited waves are often discussed in terms
		of magnetoelastic spin waves. To distinguish between traditional spin waves
		and the elastically excited spin waves we refer to them as elastic or elastic
		magnetization waves. Acoustic waves would also be a possible term.}
	
	\bibitem{Cherepov2014}
	\bibinfo{author}{Cherepov, S.} \emph{et~al.}
	\newblock \bibinfo{journal}{\bibinfo{title}{Electric-field-induced spin wave
			generation using multiferroic magnetoelectric cells}}.
	\newblock {\emph{\JournalTitle{Applied Physics Letters}}}
	\textbf{\bibinfo{volume}{104}}, \bibinfo{pages}{082403}
	(\bibinfo{year}{2014}).
	
	\bibitem{Lord1967}
	\bibinfo{author}{Lord, A.~E.}
	\newblock \bibinfo{journal}{\bibinfo{title}{Elastic wave radiation from
			simply-vibrating 180$^{\circ}$ magnetic domain walls}}.
	\newblock {\emph{\JournalTitle{Acta Acustica united with Acustica}}}
	\textbf{\bibinfo{volume}{18}}, \bibinfo{pages}{187--192}
	(\bibinfo{year}{1967}).
	
	\bibitem{Foerster2017}
	\bibinfo{author}{Foerster, M.} \emph{et~al.}
	\newblock \bibinfo{journal}{\bibinfo{title}{Direct imaging of delayed
			magneto-dynamic modes induced by surface acoustic waves}}.
	\newblock {\emph{\JournalTitle{Nature Communications}}}
	\textbf{\bibinfo{volume}{8}}, \bibinfo{pages}{407} (\bibinfo{year}{2017}).
	
	\bibitem{Weiler2011}
	\bibinfo{author}{Weiler, M.} \emph{et~al.}
	\newblock \bibinfo{journal}{\bibinfo{title}{Elastically driven ferromagnetic
			resonance in nickel thin films}}.
	\newblock {\emph{\JournalTitle{Phys. Rev. Lett.}}}
	\textbf{\bibinfo{volume}{106}}, \bibinfo{pages}{117601}
	(\bibinfo{year}{2011}).
	
	\bibitem{Hubert2008}
	\bibinfo{author}{Hubert, A.} \& \bibinfo{author}{Sch{\"a}fer, R.}
	\newblock \emph{\bibinfo{title}{Magnetic domains: the analysis of magnetic
			microstructures}} (\bibinfo{publisher}{Springer Science \& Business Media},
	\bibinfo{year}{2008}).
	
	\bibitem{McCord2015}
	\bibinfo{author}{McCord, J.}
	\newblock \bibinfo{journal}{\bibinfo{title}{Progress in magnetic domain
			observation by advanced magneto-optical microscopy}}.
	\newblock {\emph{\JournalTitle{Journal of Physics D: Applied Physics}}}
	\textbf{\bibinfo{volume}{48}}, \bibinfo{pages}{333001}.
	
	\bibitem{Hollaender2017}
	\bibinfo{author}{Holl\"{a}nder, R.~B.}, \bibinfo{author}{M\"{u}ller, C.},
	\bibinfo{author}{Lohmann, M.}, \bibinfo{author}{Mozooni, B.} \&
	\bibinfo{author}{McCord, J.}
	\newblock \bibinfo{journal}{\bibinfo{title}{Component selection in
			time-resolved magneto-optical wide-field imaging for the investigation of
			magnetic microstructures}}.
	\newblock {\emph{\JournalTitle{Journal of Magnetism and Magnetic Materials}}}
	\textbf{\bibinfo{volume}{432}}, \bibinfo{pages}{283 -- 290}
	(\bibinfo{year}{2017}).
	
	\bibitem{Queitsch2006}
	\bibinfo{author}{Queitsch, U.} \emph{et~al.}
	\newblock \bibinfo{journal}{\bibinfo{title}{Domain wall induced modes of
			high-frequency response in ferromagnetic elements}}.
	\newblock {\emph{\JournalTitle{Journal of Applied Physics}}}
	\textbf{\bibinfo{volume}{100}}, \bibinfo{pages}{093911}
	(\bibinfo{year}{2006}).
	
	\bibitem{Hengst2014}
	\bibinfo{author}{Hengst, C.}, \bibinfo{author}{Wolf, M.},
	\bibinfo{author}{Sch\"afer, R.}, \bibinfo{author}{Schultz, L.} \&
	\bibinfo{author}{McCord, J.}
	\newblock \bibinfo{journal}{\bibinfo{title}{Acoustic-domain resonance mode in
			magnetic closure-domain structures: A probe for domain-shape characteristics
			and domain-wall transformations}}.
	\newblock {\emph{\JournalTitle{Physical Review B}}}
	\textbf{\bibinfo{volume}{89}}, \bibinfo{pages}{214412}
	(\bibinfo{year}{2014}).
	
	\bibitem{Vansteenkiste2014}
	\bibinfo{author}{Vansteenkiste, A.} \emph{et~al.}
	\newblock \bibinfo{journal}{\bibinfo{title}{The design and verification of
			mumax3}}.
	\newblock {\emph{\JournalTitle{AIP Advances}}} \textbf{\bibinfo{volume}{4}},
	\bibinfo{pages}{107133} (\bibinfo{year}{2014}).
	
	\bibitem{Schloemann1964}
	\bibinfo{author}{Schl\"{o}mann, E.}
	\newblock \bibinfo{journal}{\bibinfo{title}{Generation of spin waves in
			nonuniform magnetic fields. i. conversion of electromagnetic power into
			spin-wave power and vice versa}}.
	\newblock {\emph{\JournalTitle{Journal of Applied Physics}}}
	\textbf{\bibinfo{volume}{35}}, \bibinfo{pages}{159--166}
	(\bibinfo{year}{1964}).
	
	\bibitem{Krawczyk2017}
	\bibinfo{author}{Graczyk, P.} \& \bibinfo{author}{Krawczyk, M.}
	\newblock \bibinfo{journal}{\bibinfo{title}{Coupled-mode theory for the
			interaction between acoustic waves and spin waves in magnonic-phononic
			crystals: Propagating magnetoelastic waves}}.
	\newblock {\emph{\JournalTitle{Physical Review B}}}
	\textbf{\bibinfo{volume}{96}}, \bibinfo{pages}{024407}
	(\bibinfo{year}{2017}).
	
	\bibitem{OHandley2000}
	\bibinfo{author}{O'Handley, R.~C.}
	\newblock \emph{\bibinfo{title}{Modern Magnetic Materials: Principles and
			Applications}} (\bibinfo{publisher}{John Wiley \& Sons, Inc.},
	\bibinfo{year}{2000}).
	
	\bibitem{Silva1999}
	\bibinfo{author}{Silva, T.~J.}, \bibinfo{author}{Lee, C.~S.},
	\bibinfo{author}{Crawford, T.~M.} \& \bibinfo{author}{Rogers, C.~T.}
	\newblock \bibinfo{journal}{\bibinfo{title}{Inductive measurement of ultrafast
			magnetization dynamics in thin-film permalloy}}.
	\newblock {\emph{\JournalTitle{Journal of Applied Physics}}}
	\textbf{\bibinfo{volume}{85}}, \bibinfo{pages}{7849--7862}
	(\bibinfo{year}{1999}).
	
	\bibitem{Silverman1990}
	\bibinfo{author}{Silverman, M.~P.} \& \bibinfo{author}{Badoz, J.}
	\newblock \bibinfo{journal}{\bibinfo{title}{Light reflection from a naturally
			optically active birefringent medium}}.
	\newblock {\emph{\JournalTitle{J. Opt. Soc. Am. A}}}
	\textbf{\bibinfo{volume}{7}}, \bibinfo{pages}{1163--1173}
	(\bibinfo{year}{1990}).
	
	\bibitem{Conca2013}
	\bibinfo{author}{Conca, A.} \emph{et~al.}
	\newblock \bibinfo{journal}{\bibinfo{title}{Low spin-wave damping in amorphous
			{Co40Fe40B20} thin films}}.
	\newblock {\emph{\JournalTitle{Journal of Applied Physics}}}
	\textbf{\bibinfo{volume}{113}}, \bibinfo{pages}{213909}
	(\bibinfo{year}{2013}).
	
\end{thebibliography}
\end{document}